\documentclass{article}

\usepackage{fixltx2e}
\usepackage{amsmath}
\usepackage{amssymb}
\usepackage{graphicx}

\graphicspath{{./}} 

\newcommand{\Ref}[1]{Ref.~\cite{#1}} 
\newcommand{\Fig}[1]{Fig.~\ref{fig:#1}} 
\newcommand{\Tbl}[1]{Tbl.~\ref{tbl:#1}} 

\begin{document}

\title{Equations of state of magnesium perovskite and postperovskite:
diagnostics from ab initio simulations}
\author{Roman Belousov \and Mauro Prencipe}
\date{\today}
\maketitle

\begin{abstract}
The isothermal compression of magnesium perovskite and postperovskite is
examined through the F-f plot and the diagnostic plot of Vinet universal model
theoretically from the ab initio quantum-mechanical calculations at the hybrid
Hartree-Fock / Density Functional Theory level. A purely numerical approach,
first time applied in this paper, shows that the discrepancies largely observed
between studies on the perovskite and criticized in geophysical
applications are due to the inadequate choice of the Birch-Murnaghan equation of
state; meanwhile the Vinet model is found utterly appropriate for the mineral
and infers consistent estimations of the bulk modulus and its pressure
derivative. The diagnostics of the postperovskite suggest similar conclusions.
\end{abstract}

\section{Introduction}

The family of magnesium perovskite (Mg,Fe)SiO\textsubscript{3} (space group
\textit{Pbnm}) is widely considered as the most abundant constituent of the
lower Earth mantle (~75 Vol. \% \Ref{OganovPrice2005}). It was thoroughly
studied with various ever improving techniques due to its importance in
geophysics. The existence of post-perovskite polymorphic structure (space group
\textit{Cmcm}), recently discovered and described in
\Ref{OganovOno2004,Murakami_2004}), provided for the mineralogical reasons of
geophysical anomalies met in the $D^{\prime\prime}$-layer
(mantle-core boundary), exciting interest in precise thermodynamic properties of
the phases.

In this paper the MgSiO\textsubscript{3} members of minral families are studied.
To infer the isothermal bulk moduli and their pressure/temperature
dependence, which are essential for the interpretation of seismological data,
many efforts were devoted to establish the isothermal Equation of States (EoS)
for each mineral by theoretical and experimental means. However 2 problems, that
did not reach a satisfactory state of understanding, can be traced through the
existing publications.

At first, as noted in \Ref{RossHazen1990}, the thermodynamics asserts that the
adiabatic bulk modulus must be greater than the isothermal bulk modulus, whereas
systematically upon the comparison of measurements on the magnesium perovskite
one finds the contrary relation between them, which is hardly explained by the
precision. Indeed, the values of isothermal bulk moduli as reported by many
X-ray compression experiments usually agree quite well, e.g.
\Ref{Yagi_1982,KnittleJeanloz1987,Kudoh_1987,RossHazen1990,Mao_1991,Fiquet_1998,Saxena_1999,Fiquet_2000,Vanpeteghem_2006}.
Compared to the adiabatic bulk moduli, they differ notably from the results of
Brillouin spectroscopy and shock-wave loading techniques
(\Ref{Yeganeh-Haeri_1989,Sinogeikin_2004,Deng_2008,LiZhang2005,Mosenfelder_2009}),
that as direct measurements are expected to have a superior accuracy. The
discrepancy, once attributed to the experimental errors, requires to accept a
large uncertainty which is unsatisfactory for geophysical applications
\Ref{Vanpeteghem_2006,Deng_2008}. Moreover, the values of bulk modulus pressure
derivative stand out of the range commonly found for solids.

The second issue becomes evident when considering the theoretical works, e.g.
\Ref{Karki_2000,Matsui2000,MartonCohen2002,Tsuchiya_2004,OganovPrice2005}, which
claim generally to reproduce the experimental results. Indeed, numerically the
predicted properties are found in a remarkable agreement for the accuracy in
this class of works. Nonetheless, the major part of theoretical publications
prefer the (Rose-)Vinet model of compression (VEoS), like in
\Ref{Matsui2000,MartonCohen2002,Tsuchiya_2004,OganovPrice2005}, to the
Birch-Murnaghan EoS (BMEoS), which is almost ubiquitously adopted in the
experiments. In fact, BMEoS (\Ref{Karki_2000}) appears to perform worse on
numerical comparison.

In fact, the choice of EoS can be justified by means of the designed diagnostics
on the compression data, such as the F-f plots for BMEoS or the analogous
technique for VEoS (\Ref{Angel2000}). Strictly speaking, the diagnostics should
be applied, as usually done, to reassure that the prerequisite assumptions
implied by the suggested EoS are satisfied \Ref{Stacey_1981}. Though the most of
publications do not discuss it, a review of the so called F-f plots on different
data sources is available in \Ref{Mosenfelder_2009}, which argues for large
discrepancies between the analyzed results.

The perovskite and postperovskite are delicate subjects for isothermal
compression experiments. For these minerals are high-pressure phases, they are
very stiff and undergo very small changes of volume measured by the X-ray
diffraction techniques in response to pressure. Consequently, the experiment
requires very precise and accurate measurements and pressure loading of
extremely wide range which are both difficult to carry out. The technical
challenge is naturally accompanied by larger errors. They contribute strongly to
the observed uncertainties. Meanwhile the theoretical data do not suffer
precision losses due to the stiffness of material or ranges of pressure.

With the present work we intend to show by using theoretical ab-initio
quantum-mechanical calculations that both issues are likely to stem from the
same source. An advanced mathematical approach, that we devise, is applied first
time for the theoretical diagnostics of EoS. It reveals that BMEoS of lower
orders, as adopted regularly in the isothermal X-ray compression studies, is
inadequate model for the magnesium perovskite: the inferred parameters
overestimate the isothermal bulk modulus and undervalue the pressure derivative.
On the contrary, VEoS proves to be efficient and renders elastic properties
consistently.

Also we contribute our theoretical assessments on EoS of the magnesium
postperovskite, which is still covered much less in the literature, compared
to the vast materials on the perovskite phase. Similarly, BMEoS of lower orders
is found inappropriate. Controversially, VEoS may be insufficiently flexible to
describe the postperovskite compression, possibly overestimating the bulk
modulus.

\section{Model}

Quantum-mechanical calculations were performed with the CRYSTAL09 software
package (\Ref{Dovesi_2005,Dovesi_2009}), which implements the Hartree-Fock and
Kohn-Sham Self Consistent Field (SCF) methods (\Ref{Pisani_1988}). For the
perovskite model we used a hybrid Density Functional Theory Hamiltonian WC1LYP
that adopts 84\% of the Wu-Cohen exchange term, recently suggested in
\Ref{WuCohen2006}, and 16\% of the Hartree-Fock term, alongside the
Lee-Yang-Parr correlation term (\Ref{Lee_1988}). In the case of postperovskite
we opted for a widely applied hybrid Hamiltonian B3LYP (e.g. see in
\Ref{KochHolthausen2001}). Linear Combination of Atomic Orbitals (LCAO) was
adopted as the basis set in the form of Gaussian-type functions, with the
original parameters of \Ref{Ottonello_2009} adjusted to our models according to
the variational principle.

Initial crystal structures of magnesium perovskite (\textit{pv}) and
postperovsite (\textit{ppv}) from \Ref{Horiuchi_1987,OganovOno2004} were first
optimized in the Born-Oppenheimer approximation (athermal limit) with respect to
the static lattice energy. Then a series of optimization was conducted under
unit cell volume constraints to sample the static lattice energy as a function
of volume ($V$). As implemented in CRYSTAL09, the normal mode frequencies of
atomic vibrations were computed from the derivatives of lattice energy evaluated
numerically at the zero wave vector. The Helmholtz free energy ($A$) of
insulating solid was computed within the quasi-harmonic approximation as the sum
of static lattice energy and the thermal contribution of vibrational frequencies
(\Ref{Anderson1995}).

The calculated Helmholtz energy at the set of unit cell volumes and the
temperature of interest was then fitted to a polynomial $\sum_{j=0}^N a_j V^j$,
first, by the minimax method (MM) (\Ref{Powell1981}) and, then, by the least
squares method (LSQ). We opted for the biggest possible degree of approximating
polynomial that delivered the solution of optimization problem in the uniform
norm. The LSQ optimization was performed to validate the reliability of results
and to control the precision, because its solution should converge to that of MM
(\Ref{Walsh1935}). The approximating polynomial was used to compute the pressure
$P = -\frac{\partial A}{\partial V}$ and, subsequently, to apply the diagnostics
of BMEoS and VEoS. The orders of polynomials were 6 for \textit{pv} (over 25
points of fitting) and 5 for \textit{ppv} (over 9 points of fitting).

A probe into the effect of phonon dispersion was made only on a subset of
computed data for the \text{pv} series of calculations by employing a 2x2x2
supercell. The corrections to the calculated properties were beyond the
presented precision. Thus the phonon dispersion was ignored in the following
materials.

\section{Results and discussions}

BMEoS originates from the Eulerian finite strain theory and is a model of common
choice in the physics of minerals. Conventionally the strain is expressed as
$f = \left [(V_0 / V)^{2/3} - 1 \right ] / 2$, where $V_0$ is the volume at zero
pressure. Then BMEoS emerges in various orders from the Taylor expansion of
Helmholtz Energy in terms of the strain around the minimum at zero pressure upon
truncation of the series (\Ref{Stacey_1981}). There's a technique, commonly
called the F-f plot, that is designed for a proper choice of the approximation
level (\Ref{Angel2000}). It comprises a graphical representation of the
normalized pressure $F = \frac{P}{3 f (1 + 2 f)^{5/2}}$ on the ordinate axis
versus the strain $f$ on the abscissa. On the F-f plot BMEoS of the second (BM2),
third (BM3) and forth order (BM4) correspond to a curve family graphs of a
constant (horizontal line), a line and a parabola (quadratic polynomial),
respectively.

\begin{figure}
\includegraphics[width=1\textwidth]{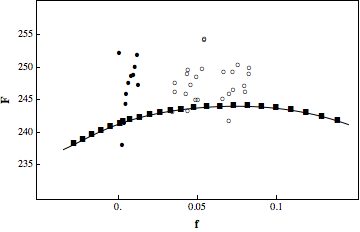}
\caption{Room temperature F-f plot for perovskite: $\blacksquare$ and solid line
- theoretical model; $\circ$ - experimental data of \Ref{Fiquet_2000}; $\bullet$
- experimental data of \Ref{Vanpeteghem_2006}.}
\label{fig:F-f_plot.room_tmpr}
\end{figure}

In \Fig{F-f_plot.room_tmpr} we compare the \textit{pv} F-f plots of our model
alongside the experimental data of \Ref{Vanpeteghem_2006,Fiquet_2000} at the
room temperature. The theoretical dependence is clearly curved with a maximum at
higher strains and varies approximately within an interval of 10 GPa. The points
of measurements are scattered in a band of about 20 GPa or 10 GPa without the
isolated outliers. The lower range of theoretical values should not surprise, as
such deviation is due to the slight overestimation of the unit cell volume,
common to quantum-mechanical models.

Legitimately, the authors of experimental works assumed a horizontal or subtly
inclined linear trend, because the ponts sprawl in an irregular manner obscuring
details of curvature. Failure of the assumption on such a large band, as the
theoretical model predicts, leads to notable biases of inferred estimations of
the elastic properties. For instance BM2EoS would overestimate the bulk modulus
at zero pressure. BM3EoS optimized for data collected under high pressure would
tend to undervalue the pressure derivative of bulk modulus, as approaching the
extremum point it decreases significantly compared to the low strains value.
Surely the experiments conducted mainly at high pressures would not fit with
those operating at low strains, as \Ref{Mosenfelder_2009} observed for the
visualized measurements of \Ref{Vanpeteghem_2006,Fiquet_2000} for instance.
Hence the higher order terms of BMEoS are critical for accurate estimations and
the equations of lower orders than BM4 are inadequate to describe the isothermal
compression on the inspected pressure intervals.

\begin{figure}
\includegraphics[width=1\textwidth]{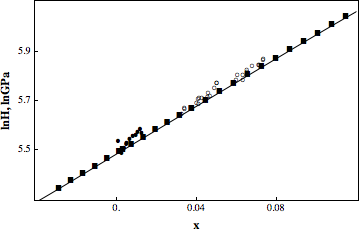}
\caption{Room temperature diagnostic plot of VEoS for perovskite: $\blacksquare$
and solid line - theoretical model; $\circ$ - experimental data of
\Ref{Fiquet_2000}; $\bullet$ - experimental data of \Ref{Vanpeteghem_2006}.}
\label{fig:Log_plot.room_tmpr}
\end{figure}

VEoS is an alternative compression model defined in terms of compression
rate $x = (V/V_0)^{1/3}$ and is admissible when the plot of points
$\ln H = \ln \left [\frac{P x^2}{3 (1 - x)} \right ]$ vs. $(1 - x)$ lie on a
straight line (\Ref{Angel2000}). This diagnostics appears missing in the
available literature, though the equation was applied to theoretical models of
perovskite, as mentioned in the introduction. The respective plot is illustrated
in \Fig{Log_plot.room_tmpr} for the same data as of \Fig{F-f_plot.room_tmpr}. It
shows that the theoretical dependence indeed forms a straight line bellow the
measurement points due to the computational accuracy. Remarkably the
experimental data are now arranged along the same line with the slope perfectly
matching the theory. We conclude thus that VEoS is an appropriate model of
compression for the perovskite.

\Tbl{blk_moduli} suggests to compare numerically the estimations of elastic
properties from various sources at the room temperature. As indicated in
\Ref{RossHazen1990}, the isothermal bulk moduli should be roughly about 5 GPa
less than the adiabatic ones. Instead, BM2EoS renders their values greater up to
10 GPa, a bit less in experiments covering a narrower range of pressures. Except
the theoretical calculation of \Ref{Karki_2000}, BM3EoS does not satisfy the
thermodynamic relation either. On the contrary, VEoS returns the isothermal bulk
moduli perfectly consistent with the adiabatic ones. This argument is very
strong, as the adiabatic moduli in \Tbl{blk_moduli} come from the direct
measurements and should in principle excel the accuracy of the results inferred
by fitting to an approximate EoS.

As anticipated in the introduction, the mineral is very stiff. Therefore a good
quality of fitting requires to explore wide ranges of pressures maintaining a
high precision of measurements. Unlikely, that the statistics of available data
is sufficient to constrain as many parameters as in BM4EoS. Even with the lesser
number of parameters, VEoS rather overestimates the pressure derivative of bulk
modulus on the measurements of \Ref{Vanpeteghem_2006}. However the value is
quite plausible once the data are extended by those of \Ref{Fiquet_2000}, while
they still agree on the bulk modulus.

Considering the level of accuracy in the class of models, our numerical results
are in a perfect agreement with the experimental estimations from VEoS. The
results obtained by LSQ and MM are practically indistinguishable. It indicates
that the solutions are convergent and well ensured within the precision of
interest. The theoretical works of other authors, who adopted VEoS, do not match
so well with the experiments though. Their procedure of parameters optimization
is not discussed in details, but such large deviation could be caused by a
sparse density of fitted points. Or one might surmise, they reduced the VEoS to
a lower degree of approximation by fixing the pressure derivative value to that
found in experiments through BMEoS. Such treatment permits to constrain the
fitted parameters better. Nonetheless VEoS is quite sensitive to $K_0^\prime$
and the error of 0.2-0.4 in the derivative might cause the observed discrepancy.

\begin{figure*}
\includegraphics[width=1\textwidth]{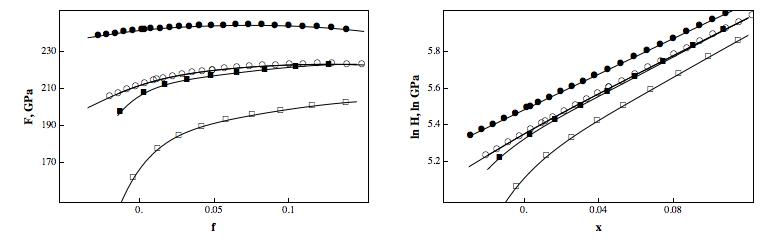}
\caption{Diagnostic plots for BMEoS (left) and VEoS (right):
$\bullet$ - \textit{pv} at 300 K;
$\circ$ - \textit{pv} at 1000 K;
$\blacksquare$ - \textit{ppv} at 300 K;
$\square$ - \textit{ppv} at 1000 K.}
\label{fig:diagnos}
\end{figure*}

\Fig{diagnos} presents the diagnostic plots for our models of \textit{pv} and
\textit{ppv} together at the room and high temperatures. The F-f plots remain
strongly curved and evidence the importance of 4th order term in BMEoS. Hence
the interpretation of diagnostics for BMEoS pertains to the high temperature
case, as well as to the postperovskite phase. The compression behavior of
\textit{pv} at the high temperature matches the model of VEoS, while the
\textit{ppv} phase reveals some curvature at the low strains. The curvature
could be due to the low density of points around the 0 pressure: there's just
one point at the negative strain and, indeed,the results of LSQ and MM
optimization were subtly divergent in that region, which is important for the
estimation of $V_0$ and hence the compression rate. Still, VEoS may overestimate
the bulk modulus and underestimate the pressure derivative at 0 strain, so long
as the curvature is not a numerical artifact. Therefore the postperovskite would
possibly require a more flexible model of compression for the accurate
estimations.

The postperovskite is unstable at ambient conditions, therefore the measurements
are conducted at elevated temperatures and pressures. Then they are extrapolated
to infer the 0 pressure and room temperature properties. Consequently, such
estimations suffer large inaccuracies. BM3EoS seems to overestimate the bulk
modulus by around 15-25 GPa (\Ref{Ono_2006,Guignot_2007,Komabayashi_2008}) with
respect to ours (206 GPa) at ambient conditions. The theoretical works that
adopt VEoS (\Ref{Tsuchiya_2004,OganovOno2004}) deliver results in the same range
as experiments. Judging from our diagnostic plots, we believe, their values are
possibly overestimated, though the hybrid Hamiltonian we used is known to
underestimate the bulk modulus.

\section{Conclusion}
Our theoretical calculations demonstrated, that BMEoS up to the 3rd order fails
to describe the compression of magnesium perovskite and postperovskite on the
pressure ranges of geophysical interest. The discrepancy observed in
\Ref{Mosenfelder_2009} between the measurements conducted at high pressures,
e.g. \Ref{Fiquet_2000}, and at low pressures (\Ref{Vanpeteghem_2006} and
others), are most likely due to the chosen EoS. It appears to cause the
systematic overestimation of the isothermal bulk modulus and the underestimation
of its pressure derivative, as verified by the direct measurements of adiabatic
bulk modulus. On the other hand, BM4EoS seems unpractical for experimental
works, because the present statistics seems insufficient to constrain so many
parameters of optimization.

However, our diagnostics reveals an excellent agreement between the Vinet model
and the compression of perovskite. Indeed, the bulk moduli inferred from VEoS
are perfectly consistent with the reported adiabatic bulk moduli. The
experiments that are found to mismatch in the review of F-f plots by
\Ref{Mosenfelder_2009} demonstrate a perfect concordance in the diagnostic plot
of Vinet model. Therewith the experiments are advised to adopt VEoS to improve
the accuracy of their inference.

The isothermal compression of our postperovskite model deviates from VEoS in the
vicinity of zero pressure. We expressed a doubt whether it's a computational
artifact due to the low density of fitted points we set in that region. Is our
diagnostic result true, the existed so far values of bulk moduli are perchance
overestimated. The theoretical computations it is recommended to follow purely
numerical methods like in this work, because they do not suffer the precision
issues related to the measurements and they are robust against failures of the
phenomenological models.
 
\bibliographystyle{spmpsci}
\bibliography{ref}

\newpage

\begin{table}
\caption{Isothermal and adiabatic bulk moduli: $K_0$ and $K_0^\prime$ are the
bulk modulus and its pressure derivative at 0 pressure; * denotes adiabatic
properties; a) theoretical calculations; b) calculations performed in this work
with the dataset of \Ref{Vanpeteghem_2006}; c) calculations performed in this
work with the combined data of \Ref{Vanpeteghem_2006} and \Ref{Fiquet_2000}.}
\label{tbl:blk_moduli}
\begin{tabular}{lrrr}
\hline\hline\noalign{\smallskip}
Reference & $K_0$, GPa & \multicolumn{1}{c}{$K_0^\prime$} & \multicolumn{1}{c}{$V_0$, \AA$^3$} \\
\noalign{\smallskip}\hline\noalign{\smallskip}
\multicolumn{4}{c}{BM EoS, 2nd order ($K_0^\prime$ is fixed to 4)} \\
\noalign{\smallskip}\hline\noalign{\smallskip}
\Ref{RossHazen1990}    & 254 &   &162.3  \\
\Ref{Yagi_1982}        & 258 &   &162.8  \\
\Ref{Kudoh_1987}       & 247 &   &162.35 \\
\Ref{Mao_1991}         & 261 &   &162.5  \\
\Ref{Fiquet_1998}      & 256 &   &162.3  \\
\Ref{Saxena_1999}      & 261 &   &162.4  \\
\Ref{Vanpeteghem_2006} & 253 &   &162.5  \\
\noalign{\smallskip}\hline\noalign{\smallskip}
\multicolumn{4}{c}{BM EoS, 3rd order} \\
\noalign{\smallskip}\hline\noalign{\smallskip}
\Ref{KnittleJeanloz1987}     & 266 & 3.9  & 162.8 \\
\Ref{Fiquet_2000}            & 253 & 3.9  & 162.3 \\
\Ref{Karki_2000}$^\text{a)}$ & 247 & 3.97 & 164.1 \\
\noalign{\smallskip}\hline\noalign{\smallskip}
\multicolumn{4}{c}{Vinet EoS} \\
\noalign{\smallskip}\hline\noalign{\smallskip}
\Ref{OganovPrice2005}$^\text{a)}$ & 261 & 4.0 & 163.3 \\
\Ref{Tsuchiya_2004}$^\text{a)}$   & 248 & 3.9 & 164.1 \\
\Ref{Matsui2000}$^\text{a)}$      & 258 & 4.0 & 162.4 \\
\Ref{MartonCohen2002}$^\text{a)}$ & 273 & 3.9 & 162.5 \\
b)                                & 245 & 5.2 & 162.6 \\
c)                                & 244 & 4.2 & 162.6 \\
\noalign{\smallskip}\hline\noalign{\smallskip}
\multicolumn{4}{c}{Other methods} \\
\noalign{\smallskip}\hline\noalign{\smallskip}
\Ref{Yeganeh-Haeri_1989}* & 246.5    &          &  \\   
\Ref{Sinogeikin_2004}*    & 253      &          &  \\
\Ref{LiZhang2005}*        & 253      & 4.4      &  \\
\Ref{Deng_2008}*          & 254      & 3.9      &  \\
\Ref{Mosenfelder_2009}*   & 254-257  & 3.9-4.3  &  \\
\noalign{\smallskip}\hline\noalign{\smallskip}
\multicolumn{4}{c}{Polynomial approximation} \\
\noalign{\smallskip}\hline\noalign{\smallskip}
this work, pv, MM  & 241 & 4.2 & 166.4 \\
this work, pv, LSQ & 241 & 4.2 & 166.4 \\
\noalign{\smallskip}\hline\hline
\end{tabular}
\end{table}

\end{document}